\documentclass[12pt]{article}
\usepackage{graphicx}

\title{ Strong  interaction  and bound states in the deconfinement phase of QCD}
\author{Yu.A.Simonov\\
 State Research
Center\\Institute of Theoretical and Experimental Physics, \\
Moscow, 117218 Russia}
 \date{}
\newcommand{\beq}{\begin{eqnarray}}
 \newcommand{\eeq}{\end{eqnarray}}
\newcommand{\be}{\begin{equation}}
 \newcommand{\ee}{\end{equation}}

\def\ga{\mathrel{\mathpalette\fun >}}
\def\fun#1#2{\lower3.6pt\vbox{\baselineskip0pt\lineskip.9pt
\ialign{$\mathsurround=0pt#1\hfil ##\hfil$\crcr#2\crcr\sim\crcr}}}
\newcommand{\veX}{\mbox{\boldmath${\rm X}$}}
\newcommand{{\SD}}{\rm SD}

\newcommand{\vex}{\mbox{\boldmath${\rm x}$}}
\newcommand{\vey}{\mbox{\boldmath${\rm y}$}}
\newcommand{\ver}{\mbox{\boldmath${\rm r}$}}

\newcommand{\vep}{\mbox{\boldmath${\rm p}$}}

\newcommand{\lan}{\langle}
\newcommand{\ran}{\rangle}
\begin{document}
\maketitle

\begin{abstract}
 Recent striking lattice results on  strong interaction and bound states above  $T_c$  can be
 explained  by the  nonperturbative
 $Q\bar Q$  potential, predicted  more than a decade ago in the framework of the field correlator method.
Explicit expressions and quantitative estimates are given for
Polyakov  loop correlators in comparison with lattice data.  New
theoretical predictions for glueballs and baryons above $T_c$ are
given.
 \end{abstract}

\section{Introduction}

There is a growing understanding nowadays that nonperturbative
dynamics plays important role in  the deconfinement phase,  for
reviews and  references see \cite{1}.

An additional part of this understanding, not contained in
\cite{1}, is the realization of the fact, that at $T_c<T<2T_c$,
the colormagnetic fields are as strong as they are in the
confinement phase, (where colormagnetic and colorelectric fields
are of the same order)  and  become even stronger above $2T_c$.
More than a decade ago the author  has argued \cite{2,3} that the
deconfinement phase transition is the transition from the
color-electric confinement phase to the colormagnetic phase,
confining in 3d. This observation was supported theoretically by
the calculation of $T_c$ \cite{2,3} and on the lattice by the
calculation of the spacial string tension at $T>T_c$ \cite{4}.

In 1991 the author has found \cite{5} that colorelectric fields
also survive the deconfinement transition in the form of potential
$V_1(r)$, which can support $Q\bar Q $ bound states in some
temperature interval $T_c<T<T_D$, while quarks acquire self-energy
parts equal to $\frac12 V_1(\infty).$ As will be shown below this
predicted picture is fully supported by recent numerous  lattice
experiments \cite{6}-\cite{13} (for a review see \cite{10}), where
$Q\bar Q$ bound states have been discovered. At the same time the
evidence for  $V_1(r)$ and selfenergies $\frac12 V_1(\infty)$ has
been also obtained on the lattice in the form of Polyakov loop
averages and of  free and internal energies above $T_c$
\cite{10,11,12}. The light quark $(m_q\sim m_s) q \bar q$ bound
states have also been observed in  \cite{13}.

The theory used in \cite{5} and below is  based on the
 powerful Method of Field Correlators MFC  \cite{14}, (for a review see \cite{15}),
 where the basic dynamic ingredients  are the field correlators
 $\lan tr F_{\mu_1\nu_1} (x_1) ... F_{\mu_n\nu_n} (x_n)
 \ran$\footnote{we omit for simplicity parallel transporters
 $\Phi(x_i, x_j)$ necessary to maintain gauge invariance, see
 \cite{15}
 for details.}.  It was shown later \cite{16} that the lowest quadratic
 (so-called Gaussian) correlator explains more than 90\% of all
 dynamics and it will be considered in what follows. The
 quadratric correlator consists of two scalar form-factors, $D$
 and $D_1$:
 \be\begin{array}{c}
 D_{\mu\nu, \lambda\sigma}(x, 0)\equiv \frac{g^2}{N_c} tr \lan
 F_{\mu\nu} (x) \Phi(x,0) F_{\lambda\sigma} (0)\ran =\\ =
 (\delta_{\mu\lambda} \delta_{\nu\sigma} - \delta_{\mu\sigma}
 \delta_{\nu\lambda}) D(x) +\frac12 \left[
 \frac{\partial}{\partial x_\mu} (x_\lambda \delta_{\nu\sigma}
 -x_\sigma \delta_{\nu\lambda}) + (\mu\lambda\leftrightarrow \nu
 \sigma)\right] D_1(x) \end{array}\label{1i}\ee
 which produce the following static potential \cite{17} between heavy quarks
 at zero temperature (obtained from the Wilson loop $r\times t$
 with $t\to \infty)$
 \be
 \begin{array}{c}
  V(r) =2r \int^r_0 d\lambda \int^\infty_0 d\tau
  D(\sqrt{\lambda^2+\tau^2}) + \int^r_0 \lambda d\lambda
  \int^\infty_0 d\tau  [-2 D(\sqrt{\lambda^2+\tau^2})+\\ +D_1
  (\sqrt{\lambda^2+\tau^2})]
  =V_D(r) +V_1(r).\end{array}
  \label{2i}\ee

One can notice that linear confinement part of potential,
$V_D=\sigma R$, is due to  correlator $D(x)$, $ \sigma = \frac12
\int D(x)d^2 x.$

At $T>0$ one should distinguish between electric and magnetic
correlators, $D^E(x), D_1^E(x)$, and $D^H(x), D^H_1(x)$ and
correspondingly between $\sigma^{(E)}$ and $\sigma^{(H)}$. It was
argued in \cite{2,3} that the principle of minimality of free
energy requires $D^E$ and electric confinement , $\sigma^{(E)},$
to vanish, while colormagnetic  correlators, $D^H(x), D_1^H(x)$
should  stay roughly  unchanged at least up  to $2T_c$.

Several years later in detailed studies on the lattice in
\cite{18} these statements have been confirmed, and indeed
magnetic correlators do not change at $1.5 T_c>T>T_c$ while
$D^E(x)$ vanishes in vicinity of $T_c$.

Not much was said about the second electric correlator $D^E_1
(x)$, since in the parametrization of \cite{18} it was found to be
 smaller than  $D^E(x)$ and hence not so important at $T<T_c$.

Meanwhile a lot of information was being accumulated on the
lattice. First of all, the Polyakov loop averages already imply
the presence of strong electric fields above $T_c$, and the main
point is  that those can  not be reduced to the perturbative
electric and magnetic, (see the  analysis in \cite{11,12}).

Recently a detailed analysis of Polyakov loop correlators was done
by the Bielefeld group \cite{9}-\cite{12}  and the singlet free
energy $F_1(r,T)$ and internal energy $U_1(r,T)$  were calculated
at $T<T_c$ and at $T>T_c$. In the latter case $F_1(r,T)$ was found
to saturate at large $r$ at the values of the order of  several
hundred MeV (e.g. for $T=1.2 T_c$ the value of $F_1(\infty, T)$
found in \cite{12} is around $0.7 \sqrt{\sigma}$, while the
internal energy  is around $3T_c$)
 and this fact cannot be explained by perturbative contributions
alone -- we consider it as the most striking  revelation of
nonperturbative electric fields above $T_c$.

 At the same time  several groups have calculated the so-called
spectral function of heavy \cite{6}-\cite{10} and light
quarkonia
  \cite{13} at $T>T_c$. In all cases sharp peaks have been observed,
corresponding to the ground state levels of $c\bar c$ at $L=0,1$
and of light $(m_q\approx m_s)$ quarkonia in $V,A,S,PS$ channels.
In both heavy and light cases the peaks are possibly displaced as
compared to $T=0$ positions  and apparently almost degenerate in
different $n\bar n$ channels.

All these facts cannot be explained  in the framework of the
commonly accepted perturbative quark-gluon plasma and call for a
new understanding of the nonperturbative physics at $T>T_c$. In
what follows we shall argue following \cite{5} that at $T>T_c$ not
only nonperturbative magnetic fields, but also strong
nonperturbative electric fields are present, which can be
calculated in MFC and explain the observed data.

\section{Dynamics of  Polyakov loops and the correlator $D_1$}

 In this section we  consider the Polyakov loop and apply to it the
nonabelian Stokes theorem and Gaussian approximation, taking first
the loop as a circle on the plane and making limiting process with
the cone surface $S$ inside loop and finally transforming cone
into the cylinder by tending the vertex of the cone to
infinity\footnote{In doing so one is changing topology of the
surface and as a result  loses the $Z(N)$ subgroup of $SU(N)$.
This however does not influence our results as long as one is
remaining in the $j=0$ sector of $Z(N)$  broken vacua (see last
ref. of \cite{1} for more discussion of $Z(N)$)}. In doing so we
are writing the nonabelian Stokes theorem and cluster expansion
for the surface $S$ which is transformed  from the  cone to the
(half)  cylinder  surface. As a result one has for the Polyakov
loop average
\be
\begin{array}{c} L= \frac{1}{N_c} tr P\exp \left(
-\frac{1}{2}\int_S d\sigma_{\mu\nu} (u) \int_S
d\sigma_{\rho\lambda}(v) D_{\mu\nu, \rho\lambda}(u,v)\right)=\\
=\exp \left\{ -\frac14 \int^{1/T}_0 d\tau\int^{1/T}_0
d\tau'\int^{\infty}_0 \xi d\xi D_1 (\sqrt{\xi^2+
(\tau-\tau')^2})\right\} .\end{array}\label{3i}\ee

In obtaining (\ref{3i}) we have  omitted the contribution of
$D(x)$ in $ D_{\mu\nu,\rho\lambda}$, since this  would cause
vanishing of $L$ in the limiting process described  above due to
the  infinite cone surface $S$. This  exactly corresponds to
vanishing of $L$ in the confinement region, observed on the
lattice. Therefore the result (\ref{3i}) refers to the
deconfinement phase, $T>T_c$.

As it is known from lattice \cite{18} and analytic calculations,
$D_1(x)$  \cite{19} exponentially falls  off at large $x$ as $\exp
(-M_1x)$, with $M_1\ga 1$ GeV and for $T\ll M_1$ one can
approximate  (\ref{3i}) as follows\footnote{The correlators
$D,D_1$ in (\ref{3i}) in principle should be taken in the periodic
form, as was suggested in \cite{21}. However for $T\leq 2 T_c$
this modification brings additional terms of the order  of $\exp
(-M_1/T)$ which are neglected below.}
\be
\begin{array}{c}
L=\exp \left( -\frac12 \int^{1/T}_0 d\nu \left(
\frac{1}{T}-\nu\right) \int^\infty_0 \xi d\xi
D_1(\sqrt{\xi^2+\nu^2})\right)\approx\\ \approx \exp \left(
-\frac{1}{2T} V_1(\infty)\right),~~ V_1(r,T)= \int^{1/T}_0 d\nu
(1-\nu T)\int^r_0 \xi d\xi
D_1(\sqrt{\xi^2+\nu^2})\end{array}.\label{4i}\ee

We turn now to the correlator of Polyakov loops following
notations from \cite{11}. Using the same limiting procedure as for
one Polyakov loop, one can apply it to the correlator $\lan
\tilde{ tr} L_{\vex}  \tilde{tr} L^+_{\vey}\ran \equiv
P(\vex-\vey)$, $\tilde{tr}=\frac{1}{N_c} tr$, representing the
loops $L_{\vex}$ and $L_{\vey}$ as two  concentric loops on the
cylinder separated by the distance $|\vex-\vey|$ along its axis,
the cylinder obtained in the limiting procedure from the cone with
the vertex tending to infinity. One can apply in this situation
the same formalism as was used in \cite{20} for  the case of the
vacuum average of two Wilson loops. For opposite orientation of
loops  using Eqs. ({21}-{28}) from \cite{20} one arrives at the
familiar form found in \cite{McLerran}
\be
P(\vex-\vey) =\frac{1}{N^2_c} \exp \left(-\frac{\tilde
F_1(r,T)}{T}\right) +\frac{N^2_c-1}{N^2_c}\exp \left(-\frac{\tilde
F_8(r,t)}{T}\right),~~ r\equiv |\vex-\vey|.\label{5i}\ee
 In the Appendix  two different ways of
derivation  of Eq.(\ref{5i}) are given, with the result \be \tilde
F_1(r,T) = V_1(r,T) + V_D(r,T), \label{6i}\ee
\be
\exp(-\tilde F_8(r,T)/T) = L_{adj} (T) \exp \{-(V_D(r,T) -\frac18
V_1 (r,T))/T\} .\label{7i}\ee

Here \be L_{adj} =\exp (-(\frac94 V_D (r^*)+\frac98 V_1
(\infty,T))/T)\label{8i}\ee is the vacuum average of the adjoint
Polyakov loop, which vanishes in the leading approximation in the
confinement phase, as it is explained in the Appendix, and nonzero
when gluon loops are taken into account, in which case $\frac94
V_D(r^*,T)\sim M_1$ and $L_{adj} \sim \exp (-M_1/T) \ll 1$.

 The suppression
of $ \exp (-\tilde F_8/T)$ in our approach in the confinement
phase has  thus the same origin as the  strong damping of the
adjoint Polyakov loop in that phase \cite{Ilgenfritz} and the
persistence of the Casimir scaling for adjoint static potential in
the interval $0\leq r < 1.2$ fm  (see \cite{117} for discussion
and references).

It is clear that  in the deconfinement phase with $D\equiv 0, ~~
V_D\equiv 0$  one has only $V_1(r)$ in both $\tilde F_1$ and
$\tilde F_8$, and all these quantities  are finite (after the
renormalization of  the perturbative divergencies specific for the
fixed contours, which are  discussed in section 3).
 Thus in the
deconfined phase one can write
\be
P(\vex-\vey) \equiv e^{-\tilde F_{q\bar q}/T}=\frac{1}{9}
e^{-V_1(r)/T} + \frac{8}{9} e^{-\left( \frac98 V_1(\infty)
-\frac18 V_1 (r)\right)/T}\label{9i}\ee where $V_1(r)$ and
$V_1(\infty)$ are renormalized. It is clear from (\ref{9i}) that
at small $r$ one has $\lim_{r\to 0} (\tilde F_{\bar q q}
(r)-V_1(r))\to T ln N^2_c$ as was noticed and measured in
\cite{11}.

 At this point one should stress the difference between
the genuine free energy $F_i(r,T), i =1.8$, which is measured with
some  accuracy on the lattice, and the  calculated above $\tilde
F_i(r, T)$. It is clear that $\tilde F_i$ do not contain the
contribution due to excitation of $Q\bar Q$ and gluon degrees of
freedom existing at finite $T$. The  latter is contained in  the
free energy $F_1(r,T)$  and in the internal energy, which we
denote $U_i(r,T) = F_i +ST$ to distinguish from our $V_i(r,T)$,
since they are not equal.

In general for nonzero temperature and  comparing to the lattice
data on heavy-quark potential one should have in mind, that
temperature effects might be of two kinds. First, the intrinsic
temperature  dependence due to changing of the vacuum structure
and the vacuum correlators and hence of our potentials $\tilde
F(r,T)$. Second, the physical quantities like $F_i(r,T), U_i(r,T)$
 are thermal averages over all excited states, e.g.
 \be
e^{-F_1(r,T)/T} = \sum_n e^{-E_n(r,T)/T}\label{dd1}
 \ee
 \be U_1(r,T) =\sum_n E_n e^{-E_n(r,T)/T}\label{dd2}
 \ee
 One can associate $\tilde F_1(r,T)=E_0(r,T)$, while the structure
 of excited spectrum can be traced in the temperature dependence
 of $F_1$ and $U_1$. E.g. assuming in the confinement phase the
 string-like spectrum and multiplicity for the multihybrid
 spectrum with two static quarks, $E_n=\sigma r+ \pi n/r, n=1,2,..$
 and multiplicity $\rho(m) =\exp (m/m_0)\theta (m-m_1), m=\pi
 n/r$, one arrives at \be F_1(r,T) = \sigma r+m_1 (1-T/m_0)- T\ln
\left(\frac{1}{T}-\frac{1}{m_0}\right)\label{dd3}\ee \be U_1(r,T)
=\sigma r   + m_1 + T/(1-T/m_0)\label{dd4}\ee The increase of
$U_1(r,T)$ below $T_c$ in the quenched case was indeed observed in
lattice calculations (see Fig.3 of \cite{12}).

 Above $T_c$ one can
see  in lattice data \cite{10} the striking drop of entropy
$S_1(\infty, T)$ and $U_1(\infty,T)$ in the region $T_c\leq T\leq
1.2 T_c$ which can be possibly explained again by the multihybrid
states occurring due to the potential $V_1(r,T)= \tilde F_1(r,T)$
connecting quarks and gluons,  and assuming that the magnitude of
$V_1(r,T)$ decreases with temperature passing at $T\approx 1.05
T_c$ the critical value enabling to bind those states of high
multiplicity. In this way one assumes that both below and above
$T_c$ in quenched and unquenched cases the dominant (in entropy)
configuration is the gluon chain connecting $Q$ and $\bar Q$ with
gluons bound together by confining string (below $T_c$) and
potential $V_1$  (above $T_c$). Thus the comparison to the lattice
data on $F_1, U_1$ needs the exact knowledge of the spectrum. In
what follows we shall associate our $\tilde F_1(r,\tau)$ with the
free energy $F_1(r,T)$, since its temperature dependence is not so
steep as that of  $U_1(r,T)$ in this region   and this discussion
will be of qualitative character, leading detailed discussion of
the spectrum to future publications.

\section{Properties of $D_1(x)$ and $F_1(r, T)$}

The correlator $D_1(x)$   was measured on the lattice \cite{18}
both below and above $T_c$, and decays exponentially with
$M_1\approx 1\div 1.5$ GeV (in the quenched case). At the same
time $D_1(x)$ can be connected to the gluelump Green's function,
and the corresponding $M_1$ for the electric correlator $D_1^E(x)$
is $M_1\approx 1.5$ GeV at zero $T$ \cite{22}. Moreover in a
recent paper \cite{19} $D_1^E(x)$ was found analytically for
$T\leq T_c$, and can be  represented symbolically  as a sum, with
perturbative part acting at small $x$,  \be D_1(x) =
D_1^{(pert)}(x) + D_1^{(np)}(x), ~~ D_1^{pert}
=\frac{4C_2\alpha_s}{\pi x^4}  + O(\alpha_s^2), \label{10i}\ee and
the  nonperturbative part  having the asymptotic form
\be
 D_1^{(np)}
(x) = \frac{A_1}{|x|}
e^{-M_1|x|}+O(\alpha_s^2),~A_1=2C_2\alpha_s\sigma_{adj} M_1,~
x\geq 1/M_1.\label{11i}\ee As will  be argued below, the form of
$D_1(x)$ (\ref{10i}) does not change for $T>T_c$, however the mass
$M_1$ and $A_1$ may be there different.

Using the asymptotics (\ref{11i}) in the whole $x$ region for a
qualitative estimate, one has \be \begin{array}{c} V_1^{(np)}
(r,T) =A_1\int_0^{1/T}(1-\nu T) d\nu \int^r_0 \frac{\xi d\xi
e^{-M_1\sqrt{\xi^2+\nu^2}}}{\sqrt{\xi^2+\nu^2}}=\\=\frac{A_1}{M_1}
\int_0^{1/T} (1-\nu T) d\nu [e^{-\nu
M_1}-e^{-\sqrt{r^2+\nu^2}M_1}]\\ =V_1^{np} (\infty) -
\frac{A_1}{M^2_1} \left[ K_1(M_1 r) M_1 r -\frac{T}{M_1} e^{-M_1r
} (1+M_1 r)  +O(e^{-M_1/T})\right]
\end{array}\label{12i}\ee  Finally the Polyakov loop exponent is
\be
L=\exp
\left(-\frac{V_1^{(np)}(\infty)}{2T}\right),~~V_1^{(np)}(\infty)
=\frac{A_1}{M_1^2} \left[1-\frac{T}{M_1}
(1-e^{-M_1/T})\right]\label{13i}\ee One can see from (\ref{13i})
that  $V_1^{(np)} (\infty)$ is finite and is of the order of few
hundred MeV in the interval $0\leq T\leq1.5 T_c$. At small $r$
from (\ref{12i}) $V_1^{(np)} \approx const\cdot r^2$. The total
$V_1=V_1^{(np)}+V_1^{(pert)}$, contains also perturbative
contribution at small $r$, which to the order $O(\alpha_s)$ is $$
V_1^{(pert)} (r) = \frac{2C_2\alpha_s}{\pi} \int^{1/T}_0 d\nu
(1-\nu T) \left( \frac{1}{\nu^2} -\frac{1}{\nu^2+r^2} \right) =
V_1^{(pert)} (\infty) + V_1^{(C)} (r,T)$$ \be V_1^{(C)}(r,T)=-
\frac{C_2\alpha_s}{r} f(r,T),~~ f(r,T) =1-\frac{2}{\pi} \arctan
(rT)-\frac{rT}{\pi} \ln [1+(rT)^{-2}]\label{14i}\ee

From (\ref{14i}) it is clear that $V_1^{(pert)}(\infty)$ is
divergent and should be renormalized, $V_1^{(pert)} (\infty)
\approx \frac{2C_2\alpha_s}{\pi} \left(\frac{1}{a} - T
lna\right),~~ a\to 0$.  Since the dominant divergent part is
$T$-independent and $V_1^{(np)}(r)\sim r^2$ at small $r$, one can
renormalize matching $V_1(r,T)$ with  the Coulomb interaction  at
small $r$, as it  was done in \cite{9,11} for $F_i(r,T)$.

As a result in the renormalized $V_1^{(pert)}(r,T)$ the term
$V_1^{(pert)}(\infty)$ can be put equal to zero, and we shall use
it in what follows.

 At this point we are able to compare $V_1(r,T)$ with the
lattice data for $ F_1(r,T)$  at  $T\geq T_c$. In Fig.1 we compare
the lattice  data for $F_1(r,T)$ taken from \cite{12} for $T=1.05
T_c, 1.2 T_c$ and $1.5 T_c$ with the potential $V_1(r,T)$ in the
form (\ref{12i}) parametrizing  $M_1$ and  $a(T)\equiv
\frac{A_1}{M_1}$ in it as \be M_1= const, ~~ a(T)=
a_0-c\frac{T-T_c}{T_c}, \alpha_s=0.3 \label{15i}\ee and find that
$M_1=0.69$ GeV  and $a_0=a(conf) = 2C_2 (f) \alpha_s \sigma_{adj}
\cong 0.432$ GeV$^2,~ c=0.36$ provides a good agreement  with the
data points at $1.5 T_c \geq T\geq T_c$, while $a(T)$ in
(\ref{15i}) smoothly matches at $T=T_c$ the amplitude of the
gluelump Green's function \cite{19}.
 One can  see that the behaviour of the total
$V_1(r,T)=V_1^C(r,T) + V_1^{(np)} (r,T)$ which has a Coulomb part
at smaller $r$ and saturates at $V_1=V_1^{(np)} (\infty)$ is
qualitatively very similar to the behaviour of $F_1(r,T)$  as a
function of $r$.
 We also
compare in Fig.2  our results with lattice data \cite{11} for the
Polyakov loop (\ref{13i}) and find reasonable
agreement.
 It is clear that both Fig.1
and Fig.2 are qualitative illustrations, and  for quantitative
comparison one needs knowledge of excitation spectrum and analytic
or lattice predictions for $a(T), M_1(T)$ which will be given
elsewhere \cite{Giacomo}, \cite{32}.

 \unitlength=1cm

\begin{figure}[h]
\begin{picture}(15,8)
\put(1.5,1){\includegraphics[height=7cm]{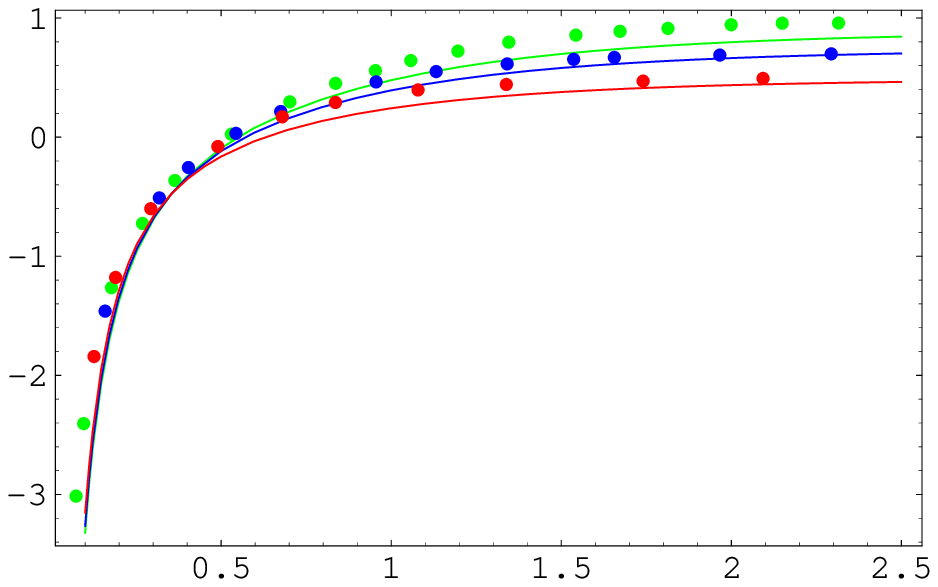}}
\put(7,0){$r \sqrt{\sigma}$}
\put(0,4){\rotatebox{90}{$F_1/\sqrt{\sigma}$}}
\end{picture}
\caption{ A comparison of behavior of $V_1(r,T)=V_1^{(np)}(r,T) +
V_1^{(C)}(r,T)$ Eqs.(16),(18),(19)(solid lines,
$T/T_c=1.05;1.2;1.5$ from above), with the singlet free energy
$F_1(r,T)$ measured in ref. \cite{12} (filled circles.)}

\end{figure}

\clearpage

\begin{figure}[htb]
\begin{picture}(15,8)
\put(1.5,1){\includegraphics[height=7cm]{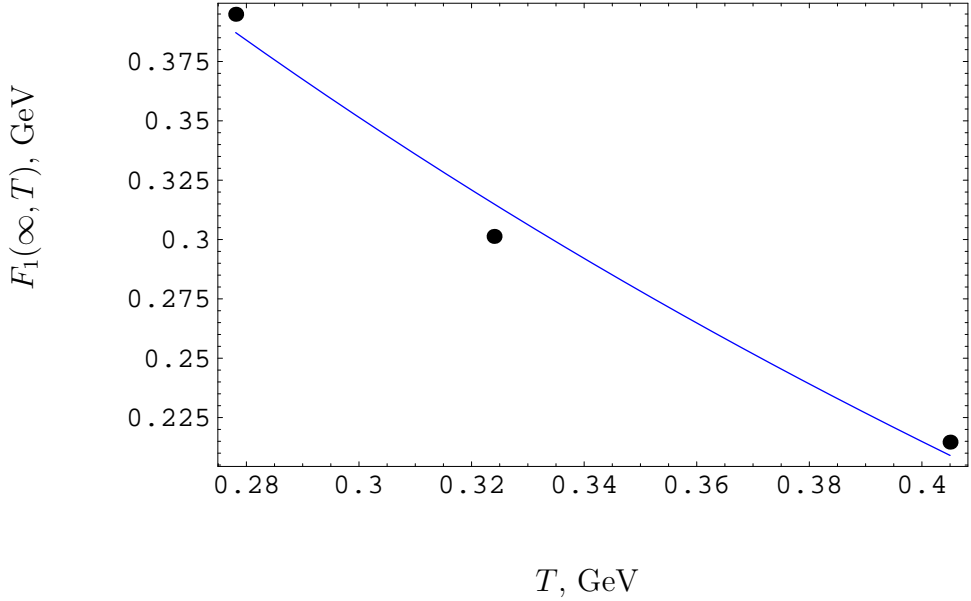}}
\put(7,0){$T$, GeV}
\put(0,4){\rotatebox{90}{$F_1 (\infty, T)$, GeV}}
\end{picture}
\caption{ The Polyakov loop exponent $V_1(\infty,T)$ as a function
of $T$ (GeV) from Eqs.(17),(19)(solid line), in comparison with
lattice data for $ F_1(\infty,T)$ in ref. \cite{11} (filled
circles).}

\end{figure}

 \section{Bound states of $q\bar q$ in the deconfinement region}

In the  recent    lattice studies sharp peaks have been found
 in the spectral function of $c\bar c$ system \cite{6}-\cite{10}, which can be
 associated with the quark-antiquark bound states  surviving at
 $T\geq
 T_c$.

 To understand qualitatively whether the interaction $V_1(r,T)$
 can support bound states, one can
  use the
 Bargmann  condition   \cite{23} for monotonic attractive potentials
 \be
 2\bar m \int^\infty_0 r dr |U(r,T)|>1\label{17i}\ee
where $2\bar m=m_c$, and $U(r,T)=V_1(r,T) - V_1(\infty, T)$ which
yields the condition for the bound $S$-states. Taking $m_c=1.4$
GeV, $M_1=0.6$ GeV, one can
 deduce that
 $V_1^{(np)}(r,T)$ can support bound states in some interval of
temperature $T_c<T<T_D$ where $T_D \sim 1.5 \div 2  T_c$ and exact
value depends on terms of the order  $O(e^{-M_1/T})$ and therefore
is beyond the scope of the present paper. This conclusion roughly
agrees with lattice calculations in \cite{6}-\cite{9} and with the
calculation done in \cite{24} and \cite{25}.

 One can consider
gluons and the $gg$ system in the same way as it was done for the
$c\bar c$ system. To this end one should first multiply $V_1(r,T)$
found earlier in (\ref{12i}-\ref{14i}) by $\frac{C_2(adj)}{C_2(f)}
=\frac94$, thus defining $V_1^{(adj)} (r,T) =\frac94 V_1(r,T)$.
This potential can be considered as the interaction kernel in the
Hamiltonian of the $gg$ system as it was done  in \cite{26}. This
Hamiltonian with the introduction of the einbein gluon mass $\mu_g
\cong \lan \sqrt{\vep^2}\ran$ has the form of the Schroedinger
equation and the condition (\ref{17i})
 can be  applied. Since $\mu_g\approx 0.6$ GeV \cite{26}, an
 additional (with respect to quarks) factor for the
 nonperturbative part in (\ref{17i}) is $\frac94 \frac{\mu_g}{m_c}
 \approx 0.96$. Hence two-gluon glueballs should be formed in
 approximately the same temperature interval as charmonium
 states.In lattice calculations \cite{27} scalar glueball was
 studied below and around $T_c$, and its width is increasing with
 $T$.
 Now we come to the baryon case and using the general formalism
 \cite{28} to represent the triple Wilson loop of trajectories of
 3 quarks and of the string junction in terms of field correlators
 $D$ and $D_1$. From  \cite{28} one has for 3
 static quarks at distances  $r_1, r_2, r_3$ from the string
 junction
 \be
 V_{3q} (\ver_1, \ver_2, \ver_3;T) = \sum^3_{i=1} V_D (r_i,T)+ \frac12 \sum_{i>j} V_1(\ver_i
-\ver_j,T)\label{19i}\ee where $V_1 (r, T)$ is given in
(\ref{12i}), (\ref{14i}).
  In  the
deconfinement phase when $r_i=R, i=1,2,3,$  one obtains $V_{3q}
=\frac32 V_1(\sqrt{3}R).$ For the perturbative part one has  from
(\ref{19i}) $V_{3q}^{(C)}=\frac12 \sum_{i>j} V_1^{(C)}(r_{ij},
T)|_{r_{ij}=R}$.

For the nonperturbative part from Eq.(\ref{19i}) it follows that
$V_{3q}^{(np)} (R\to \infty) =\frac32 V_1^{(np)} (\infty)$. One
can check that this prediction  and the general form of $V_{3q} =
\frac32 V_1(\sqrt{3}R)$ as function of $R$ is  supported by the
recent measurement of singlet free energy of the $3Q$ system in
\cite{29} at $T>T_c$. Thus it is of interest to measure the
spectral functions of baryons at $T>T_c$ in the same way as it was
done for mesons.

\section{Summary and conclusions}

Citing the 1991 paper \cite{5} when the magnitude of $D_1$ was not
exactly known "...Using an exponential parametrization for $D_1$,
we can find $D_1^{E}$ with parameter values which satisfy the
condition for the appearance of levels. In this case $\varepsilon
(r)$ (our $V_1^{(np)}(r,T)$) is a well with  a behaviour
$\varepsilon (r) \sim r^2$ as $r\to 0$ and $\varepsilon (r) \to
const >0$ as $r\to \infty$.  The quark and antiquark are thus
bound but there exists a threshold $\varepsilon(\infty)$ above
which quarks fly apart, each acquiring a nonperturbative mass
increment $\delta m =\frac12 \varepsilon (\infty)$...". In the
present paper this picture was  further substantiated and
quantified using lattice and analytic knowledge on $D_1$.
Comparing to recent lattice data in \cite{6}-\cite{13} it was
shown that this picture is qualitatively supported by data, and
new proposals have been done for searching the glueball and baryon
systems at $T>T_c$.

The results obtained on the lattice \cite{6}-\cite{13} and in the
present approach establish a new picture of the QCD thermodynamics
at $1.5 T_c> T\geq T_c$ widely discussed in \cite{24}.  As a
 new feature compared to
  the works \cite{24} the main emphasis in this paper is done on the selfenergies
  $(\frac12 V_1 (\infty) \equiv  \varepsilon_q$
  for quarks and $\frac98 V_1 (\infty) \equiv \varepsilon_g$  for gluons)
  which are large  $(V_1(\infty, T_c)> F_1 (\infty, T_c) \cong 600$ MeV for $n_f =2 $
  \cite{Kaczmarek}) and cancel each other at small distances for
  white bound states, like $q\bar q$, $(gg)_1, (qg\bar q)_1,
  (qg...g\bar q)_1$ etc.
   In contrast to that colored states are higher in potential and
   mass by several units of $\varepsilon_q$ and $\varepsilon_g$ and   are suppressed  by
   the corresponding  Boltzmann factors.
As a result in this region white bound states of quarks and gluons
are energetically preferable, while individual quarks and gluons
acquire selfenergies, so that the thermodynamics of the system
resembles that of the neutral gas, and  for higher temperature
$T>1.5 \div 2 T_c$ a smooth transitoon to the "ionised" plasma of
colored quarks and qluons possibly occurs.

This new state of the quark-gluon matter should be taken into
account when considering ion-ion collisions. For more discussion
of the thermodynamics above $T_c$ see \cite{24} and refs. therein.

It was noted before \cite{30} that the behaviour of the free and
internal energies above $T_c$, with a bump around $T\sim 1.1 \div
1.2 T_c$ in $\frac{\varepsilon -3P}{T}$ can be explained if gluons
are supplied with the nonperturbative mass term of the order of
0.6 GeV, while for higher $T$ this mass is less important. This
can be easily understood now taking into account the value of
$\frac12 V_1 (\infty, T)$ and its decreasing with growing $T$. In
this way the nonperturbative dynamics in the form of correlator
$D_1$ can explain the observed dynamics of the deconfined QCD.

 A more
detailed analysis of bound states requires  explicit calculation
of $Q\bar Q$ and $3Q$ bound states taking into account  spin
splitting in the mass of $P$-wave charmonia and quasi-degeneration
of spectra of light $q\bar q$ $V,A,S,PS$ states observed in
\cite{13}. Here spin-dependent forces are different from the
confining case, since only the correlator $D_1$ contributes, and
one can list the corresponding terms in \cite{17,31}.
 It is interesting to note, that due to the vector character
   of $V_1(r,T)$, not violating chiral symmetry, bound states of massless quarks should exhibit parity doubling.
  All this
analysis is now in progress \cite{32}.

The author is grateful to N.O.Agasian, S. M.Fedorov and
V.I.Shevchenko for discussions and D.V.Antonov, N.Brambilla,
E.-M.Ilgenfritz,  A.Vairo and C.-Y.Wong for useful correspondence,
and P.Petreczky for valuable remarks and comments.

 The work  is supported
  by the Federal Program of the Russian Ministry of industry, Science and Technology
  No.40.052.1.1.1112, and by the
grant for scientific schools NS-1774. 2003. 2.\\

\vspace{2cm}

{\bf Appendix 1}\\

{\bf  Derivation  of  the Polyakov loop correlator }\\

 \setcounter{equation}{0} \def\theequation{A1.\arabic{equation}}

 We give here two different derivations of $\tilde F_i (r, T),~~
 i=1,8$. The first  is based on the correlator of two concentric
 Wilson loops, derived in \cite{20}, in which case $\tilde
 F_{1,8}$ are expressed in terms of surface integrals of field
 correlators $D_{\mu\nu,\rho\lambda}(u,v)$
 \be
 I(S_i, S_k) \equiv \int_{S_i}  d\sigma_{\mu\nu} (u) \int_{S_k}
 d\sigma_{\rho\lambda} (v) D_{\mu\nu,\rho\lambda}
 (u,v)\label{A.1}\ee

 In this way one obtains for two oppositely directed Polyakov
 loops from Eqs. (29), (23), (20) of \cite{20}
 \be
 \tilde F_1(r,T)/T =\frac12 I(S_{12}, S_{12})\label{A.2}\ee
\be
 \tilde F_8(r,T)/T =\frac12 I(S_{1}, S_{1}) +\frac12 I(S_2,S_2) +\frac{1}{N_c^2-1} I(S_1,S_2)\label{A.3}\ee

 Here $S_1$ is the surface on the cylinder with circumference
 $1/T$ extending from the loop 1 at  coordinate $\vex$ in the
 direction $\vey$ to infinity, the surface $S_2$ is also infinite
 surface from the loop 2 at coordinate $\vey$ in the  same
 direction (the answer does not depend on the  choice of this
 direction).

 The surface $S_{12}$ lies on the cylinder between the loops 1 and
 2. Note that surface orientation in (\ref{A.1})  is fixed to be same.
 Calculation of $\tilde F_1$ according to  (\ref{A.2}) reduces to that
 of the Wilson loop and yields
 \be
 F_1(r,T) = V_1 (r,T) + V_D(r,T)\label{A.4}\ee
 where $V_1(r,T)$ is given in (\ref{4i}) and $V_D$ is
 \be
 V_D(r,T) =2\int^{1/T}_0 d\nu (1-\nu T) \int^r_0 (r-\xi) d\xi
 D(\sqrt{\xi^2+\nu^2})\label{A.5}\ee

 Calculation of $\tilde F_8$ is more subtle. To this end one can
 use connection of $D_1(x)$ to the gluelump Green's function
 $G_{\mu\nu}(x)=\delta_{\mu\nu} N_c(N_c^2-1) f(x^2)$ \cite{19}
 $D_1(x)= -\frac{2g^2}{N_c} (N^2_c-1) \frac{df(x^2)}{dx^2},$
 $f(x^2\to 0) \sim \frac{1}{4\pi^2 x^2}$
 and inserting this into (4), one has
 \be
 V_1 (r,T) = \frac{g^2(N^2_c-1)}{N_c} \int^{1/T}_0 d\nu (1-\nu T)
 [f(\nu^2) - f(r^2+\nu^2)]=
 V_1(\infty, T) + v_{ex} (r,T)\label{A.6}\ee

 One can see that $V_1(\infty,T) \equiv V_Q+V_{\bar Q}$ is the sum
 of equal selfenergy parts of $Q$ and $\bar Q$, while $v_{ex}
 (r,T)$ describes interaction due to one gluelump exchange between
 $Q$ and $\bar Q$.

 Note that $V_1(0,T)=V_1(\infty, T)+ v_{ex} (0,T)=0$ and
 $v_{ex}(\infty, T)=0$. Therefore $v_{ex}$ appears only in $I(S_1,
 S_2)$ in (\ref{A.3}) and one  should restore there the original
 (opposite) orientation of loops $L_{\vex}$ and $L_{\vey}^{+}$ to get the correct
 sign of $v_{ex}$ (the same sign and factor appears in the second derivation below).

 From (\ref{A.1}) one obtains for $I(S_i,S_i)$
 $$
 \frac12 I(S_2, S_2) = V_D (r^*) + V_{\bar Q} ,~~ \frac12
 I(S_1, S_1) =\frac12I (S_2, S_2)+ V_D(r),$$ \be
  I(S_1, S_2) =
 -v_{ex} (r,T) + I (S_2, S_2)\label{A.7}\ee

 As a result one can write using (\ref{A.3}) for  $N_c =3$
 $$
 \tilde F_8 (r,T) =\frac94 V_D (r^*, T) + V_Q + V_{\bar Q} -
 \frac18 v_{ex} (r,T) +$$ \be + V_D(r,T) =\frac98 V_1 (\infty, T)
 -\frac18 V_1 (r,T) + \frac94 V_D(r^*,T)+ V_D (r,T)\label{A.8}\ee

 In (\ref{A.8}) the value of $r^*$ is infinitely large, when one
 neglects the valence gluon loops, as it is done everywhere above.
 In this case $V_D(r^*,T)\to \infty$ and the term $\exp (-\tilde
 F_8/T)$ vanishes in the confinement region.
  This is in line with the strong damping of the adjoint Polyakov
  loop in this region observed  on the lattice \cite{Ilgenfritz}, and with
  the persistence of Casimir scaling for adjoint static potential
  for $0\leq r \leq 1.2$ fm found on the lattice (see discussion
  and refs.in \cite{117}.

  The correction due to the gluon determinant, producing
  additional gluon loops becomes important for $r\geq 1.2$ fm (see
  discussion in the second ref. in \cite{117}) and makes finite the
  value of $V_D(r^*,T) \approx \sigma r^*,  r^* \approx 1.2$ fm/2 =0.6 fm.
  $$
  \exp (-\tilde F_8 (r,T)/T) = L_{adj} (T) \exp \left(-\frac{V_D(r,T)
  -\frac18 V_1(r,T)}{T}\right),
$$ \be L_{adj} (T) =\exp (-\frac98 V_1 (\infty,T) -\frac94 V_D
(r^*,T)).\label{A.9}\ee

In (\ref{A.9}) the effects of loop-loop interaction and of the
total (adjoint) loop are separated. Physically the result
(\ref{A.9}) can be easily understood: in  absence of the internal
interaction one has do  with the adjoint Polyakov loop, which
strongly changes around $T_c$, namely $L_{adj}(T<T_c)$   is much
smaller than $L_{adj}(T>T_c)$.

The behaviour similar to (\ref{A.9})  was observed on the lattice,
see Fig. 3 of  second ref. \cite{10} and ref.\cite{Kaczmarek}.
Here one observes linear growth in $r$ of $\tilde F_8(r,T)$ below
$T_c$, and repulsive Coulomb behaviour from $\frac18 V_1(r,T)=
\frac{\alpha_s}{6 r}$.

In the second  derivation one is   connecting two loops using
parallel transporters and using  the  completeness relation \be
\delta_{\alpha_1\beta_1} \delta_{\alpha_2\beta_2} =\frac{1}{N_c}
\delta_{\alpha_1\beta_2} \delta_{\beta_1\alpha_2} + 2
t^a_{\beta_2\alpha_1}t^a_{\beta_1\alpha_2}\label{A.10}\ee so that
the second term produces the adjoint Wilson loop on the cylinder
surface. $$\exp (-\tilde F_8/T) = 2 tr \lan (U(\vex, \veX;0) t^a
U(\veX,\vey, 0) L(\vey)\times $$
\be
\times U(\vey, \veX,t) t^aU(\veX,\vex,t)
L^+(\vex)\ran\label{A.11}\ee

This  can be compared to the approach, suggested in \cite{37}. Our
results however are different from those of \cite{37} in that both
perturbative and nonperturbative interactions in $\tilde F_1$ and
$\tilde F_8$ are different (and calculable through $D_1$). Now
(\ref{A.11}) can be rewritten using cluster expansion and
nonabelian Stokes theorem, which finally results in the same
equations as in (\ref{A.9}). Alternatively one can use the technic
exploited in \cite{38} to separate the contributions of
perturbative exchanges from the nonperturbative confining terms.

For the first ones one commutes as in \cite{38} the color
generators $t^c$ of exchanged gluon (gluelump) with  $t^a$
(\ref{A.11}) acording to the equality $t^ct^at^c=-t^a/2N_c$ which
finally gives in (\ref{A.11}) the adjoint Coulomb interaction
$\frac{\alpha_s}{6 r} \to - \frac18 V_1(r)$, while the selfenergy
parts and confining terms arise from sequences $t^ct^{c'} t^a\to
\delta_{cc'}t^a$ and do not change sign. In this way one arrives
at the same answer as given in (\ref{A.9}).


\begin{thebibliography}{99}

\bibitem{1}
P.Arnold, L.G.Yaffe,  Phys. Rev. {\bf D52} (1995) 7208,
hep-ph/9508280; L.G.Yaffe, Nucl. Phys. Proc. Suppl. {\bf 106},
(2002) 117, hep-th/0111058;\\ R.D.Pisarski, hep-ph/0203271.

\bibitem{2} Yu.A.Simonov, Pis'ma
 v JETF, {\bf 58} (1995)  357; Phys. At. Nucl. {\bf 58} (1995) 309,  hep-ph/9311216.

 \bibitem{3}Yu.A.Simonov, Lecture at the International School of
     Physics "Enrico Fermi", Varenna, 27 June--7 July 1995, "Varenna 1995, selected topics in
nonperturbative QCD", 319-337; hep-ph/9509404.
\bibitem{4}
C.Borgs, Nucl. Phys. {\bf B261} (1985) 455; E.Manousakis and
J.Polonyi, Phys. Rev. Lett. {\bf 58} (1987) 847; G.S.Bali et al.
Phys. Rev. Lett. {\bf 71} (1993)  3059; G.Boyd, J.Engels, F.Karsch
et al., Nucl. Phys. {\bf B469} (1996) 419.

\bibitem{5} Yu.A.Simonov, JETP Lett. {\bf 54} (1991) 249.

\bibitem{6} I.Wetzorke, F.Karsch, E.Laermann, P.Petreczky and
S.Stickan, Nucl. Phys. Proc. Suppl. {\bf 106} (2002) 510,
hep-lat/0110132.

\bibitem{7} T.Umeda, K.Nomura and H.Matsufuru, Eur. Phys. J.C. (2004), hep-lat/0211003.

\bibitem{8} M.Asakawa and  T.Hatsuda,  Phys. Rev. Lett. {\bf 92}
(2004) 012001.

\bibitem{9} S.Datta, F.Karsch, P.Petreczky and I.Wetzorke,   hep-lat/0208012; Phys.
Rev. {\bf D69} (2004) 094507;\\  F.Karsch et al. Nucl. Phys. {\bf
A715} (2003) 863;  P.Petreczky, J. Phys. {\bf G30} (2004) 431;\\
T.Umeda,  H.Matsufuru, hep-lat/0501002.

\bibitem{10}
P.Petreczky, Plenary talk at Lattice 2004, hep-lat/0409139,
{P.Petreczky, hep-lat/0502008}.
\bibitem{11}
O.Kaczmarek, F.Karsch, P.Petreczky and F.Zantow, Phys. Lett. {\bf
B543} (2002) 41, hep-lat/0207002.

\bibitem{12}
O.Kaczmarek, F.Karsch, P.Petreczky and F.Zantow, hep-lat/0309121;
hep-lat/0406036.

\bibitem{13} M.Asakawa, T.Hatsuda and Y.Nakahara, Nucl. Phys. {\bf
A715} (2003) 701, hep-lat/0208059.



\bibitem {14}
  H.G.Dosch, Phys. Lett. B {\bf 190}, (1987) 177;\\
  H.G.Dosch and Yu.A.Simonov, Phys. Lett. B {\bf 205}, (1988) 339;\\
   Yu.A.Simonov, Nucl.  Phys.  B {\bf 307}, (1988) 512.

\bibitem{15} A.Di Giacomo, H.G.Dosch, V.I.Shevchenko, Yu.A.Simonov, Phys. Rept. 372 (2002)
319.

\bibitem{16}  V.I.Shevchenko, Yu.A.Simonov, Phys.  Rev. Lett. {\bf 85} (2000) 1811.

\bibitem{17}
Yu.A.Simonov, Nucl. Phys. {\bf B324} (1989) 67.

\bibitem{18}
A.Di Giacomo and H.Panagopoulos, Phys. Lett. {\bf B285}(1992)
133;\\
 A.Di Giacomo, E.Meggiolaro and  H.Panagopoulos, Nucl. Phys.
{\bf B483}   (1997) 371;\\ M.D'Elia, A.Di Giacomo and E.Meggiolaro
Phys. Rev.   {\bf D67}  (2003) 114504;\\ G.S.Bali, N.Brambilla and
A.Vairo, Phys. Lett. {\bf B42} (1998) 265.


\bibitem{19} Yu.A.Simonov, hep-ph/0501182.

\bibitem{20}  V.I.Shevchenko, Yu.A.Simonov, Phys.  Rev.  {\bf D66} (2002) 056012.

\bibitem{McLerran}  L.G.Mc Lerran, B.Svetitsky, Phys. Rev. {\bf
D24} (1981) 450;\\ S.Nadkarni, Phys. Rev. {\bf D33} (1986), ibid
{\bf D34} (1986) 39904.

\bibitem{21} N.O.Agasian, Phys. Lett.  {\bf B562}
(2003) 257.
\bibitem{Ilgenfritz} E.-M.Ilgenfritz, private communication.

\bibitem{117}
    V.I.Shevchenko, Yu.A.Simonov, Phys. Rev. Lett. {\bf 85 } (2000)
  1811; hep-ph/0001299.

\bibitem{Giacomo} A.Di Giacomo, E.Meggiolaro, Yu.A.Simonov (in
preparation).

\bibitem{22} Yu.A.Simonov, Nucl. Phys. {\bf B592} (2001) 350.

\bibitem{23} F.Calogero, Variable Phase Approach to Potential
Scattering, Academic Press, NY, 1967.

\bibitem{24} Cheuk-Yin Wong, hep-ph/0408020;\\ E.V.Shuryak,
I.Zahed, hep-ph/0403127; \\ H.-J.Park, C.-H.Lee, G.E.Brown,
hep-ph/0503016.

\bibitem{25} A.Mocsy, P.Petreczky, hep-ph/0411262.


\bibitem{26}
A.B.Kaidalov, Yu.A.Simonov, Phys. Lett. {\bf B477} (2000) 163;
Phys. Atom. Nucl. {\bf 63} (2000) 1428, hep-ph/9911291.

\bibitem{27} N.Ishii, H.Suganuma, H.Matsufuru, Prog. Theor. Phys.
Suppl. {\bf 151} (2003) 166;\\  N.Ishii, H.Suganuma,
hep-lat/0312040.


\bibitem{28} Yu.A.Simonov, Phys. Atom. Nucl. {\bf 66} (2003) 338,
hep-ph/0205334.

\bibitem{29} K.H\"{u}bner, O.Kaczmarek, F.Karsch, O.Vogt,
hep-lat/04088031.

\bibitem{Kaczmarek} O.Kaczmarek, F.Zantow, hep-lat/05030117.

\bibitem{30} H.G.Dosch, H.J.Pirner, Yu.A.Simonov, Phys. Lett.  {\bf B34} (1995) 335;\\
N.O.Agasian, D.Ebert, E.-M.Ilgenfritz, Nucl. Phys. {\bf A637}
(1998) 135;\\
 P.Petreczky, F.Karsch, E.Laermann,  S.Stickan and I.Wetzorke, Nucl.
Phys. Proc. Suppl. {\bf 106} (2002) 513, hep-lat/0110111.

\bibitem{31}
A.M.Badalian, Yu.A.Simonov, Phys. At. Nucl.   {\bf 59} (1996)
2164.
\bibitem{32} N.O.Agasian, S.M.Fedorov and Yu.A.Simonov (in
preparation).

\bibitem{37} {O.Philipsen, Phys. Lett. {\bf B535} (2002) 138;
O.Jahn, O.Philipsen, hep-lat/0407042}

\bibitem{38} {Yu.A.Simonov, S.Titard, F.J.Yndurain, Phys. {\bf
B354} (1995) 435}
\end{thebibliography}
\end{document}